\documentstyle{qcdparis}
\begin{document}
\pagestyle{plain}
\title{Nuclear phenomena at HERA energies}
\author{Leonid Frankfurt$^{\dag}$ and Mark Strikman$^{\dagger}$}
\affil{$^{\dag}$ School of Physics and
  Astronomy Raymond and Beverly Sackler Faculty of Exact Sciences,\\Tel Aviv
University,
Israel\\
$^{\dagger}$  Physics Department, Pennsylvania State University
University Park, PA 16802, U.S.A.  }

\abstract{We argue that current data on nuclear shadowing confirm expectations
of the color singlet models of nuclear shadowing.
We demonstrate that unitarity constrains require nuclear shadowing
for gluons to be very significant already for $x \le 10^{-3}$ . Physics of
coherent diffraction production of vector mesons off light nuclei is
also discussed.}

\resume{Nous montrons que les donnees actuelles concernant l'effet d'ecrantage
nucleaire confirment les previsionns des modeles de "couleur singlet"
d'ecrantage nucleaire. Nous demontrons que les contraints d'unitarite
exigent que l'effet d'ecrantage nucleaire pour les gluons doit etre deja
tres significatif pour les $x \le 10^{-3}$.
Nous discuterons aussi de la physique
de la production diffractive coherente des mesons vecteurs sur un noyau
legere.}

\twocolumn[\maketitle]
\fnm{7}{Talk given in the
nuclear  session
at the Workshop on Deep Inelastic scattering and QCD,
Paris, April 1995}

\section{Nuclear shadowing for parton distributions for $x \sim 10^{-2}$}

Use of nuclear targets in DIS provides sensitive tests of interplay of
perturbative and nonperturbative
dynamics in     small
$x$ QCD.  Currently the most revealing example  is  the  nuclear  shadowing
phenomenon. Historically two qualitatively different QCD hypotheses  were  made
about   the  dominant  source  of  nuclear   shadowing   at
small $x$. One was fusion   of  combining  partons  from  different
nucleons  \cite{MQ,Q,Close}   another   was   color   screening   mechanism
\cite{FS88,FLS,BL,NZ} in which   dominant mechanism of shadowing is soft
color singlet interaction of $q \bar q$ pairs of transverse close to
the hadron size with nucleons of the target.  The second mechanism is a QCD
extension of the parton model observation of Bjorken that  $q \bar q$
hadron-like configurations    in the virtual photon    give    a leading
twist, scaling contribution  to the deep inelastic  cross  section   -  the
so called aligned jet model \cite{Bj}.

     Two key predictions of the color singlet  mechanism are now
confirmed experimentally - NMC has finally observed \cite{scale} the pattern of
scaling
violation predicted in \cite{FLS}, diffraction in DIS expected in the color
screening mechanism \cite{Bj,FS88} {\it to be the leading twist effect}
 was observed in DIS off nuclei
\cite{NMC1}  and  at HERA.   It is worth emphasizing that in the fusion
mechanism no
leading twist diffraction is expected since no color singlet is exchanged in
$t-$ channel.

     The QCD analysis  \cite{FLS}  indicates  that  at  intermediate  $Q^2$
shadowing is present in all channels - gluons, sea and valence quarks while
the enhancement at larger $x \sim 0.1$ is present for  valence  quarks  and
gluons only. Further experimental studies  are  necessary  to  check  these
predictions.

    Current versions of the color screening  models    usually
assume that the eikonal type interactions with color singlet  $q  \bar  q$
pair dominate. The average cross section  for  this  interaction  is  about
15-18 mb for $\nu \sim 200 ~GeV$ \cite{FS88,BL}.
Fluctuations in the interaction cross section  lead  to  a  small
correction only. However   in  QCD  mechanism  of  shadowing  is  more
complicated since the interaction of small transverse
size $q \bar q$ configurations  (which  are
not screened in the eikonal approximation) is given by \cite{BBFS93}.
\begin{equation}
\sigma(b^2)={\pi^2\over 3}\left[ b^2\alpha_s(Q^2)
xG_{T}(x,Q^2))
\right]_{x=Q^2/s,  Q^2 \approx 15/b^2} \ ,
\label{1}
\end{equation}
One can see  from this equation that as soon as gluon shadowing is  present
interaction of small size $q \bar q$ configurations
is screened as well {\it in the
leading twist}.  To study interplay  of  the  contribution  of  small
     and large size configurations it is necessary  to  study  in  greater
detail diffractive processes.  We find that total  cross  section  of
diffraction is  rather insensitive  to  dispersion  of  the  interaction
strength:  two   extreme assumptions - no fluctuations in  the  interaction
cross section  and  the two-component model with configurations which are
not screened contributing $30 \%$ of $\sigma_{\gamma^*N}(x \sim 10^{-2}, Q^2
\simeq few GeV^2)$ and configurations  interacting  with $\pi  N$  cross
section  contributing $70\%$ - lead   to   practically   the   same   ratio
  of coherent diffractive and total cross sections for $x \sim 10^{-2}$. It
can be approximated as
\begin{equation} {\sigma_{coh.diffr.}
\over\sigma_{tot}}_{|200 \ge A \ge  12}  \approx  0.1
\left({A \over 10}\right)^{0.25}
\end{equation}
This value is consistent with the observations of E-665
 \cite{NMC1}.  Clearly to get a deeper insight into dynamics of shadowing one
has  to  study  the final states in a more exclusive way.

\section{Physics issues for Nuclear beams at HERA}
Currently there is discussion of studies of DIS scattering  off  nuclei  at
HERA. This could open qualitatively new opportunities for understanding  of
the small x dynamics and would have serious impact for the studies of  $AA$
collisions at LHC.

\subsection{Shadowing of hard processes at very small x}
Cross section of any deep inelastic hard process at small x can
be expressed in the leading
twist through the cross section of interaction of $q \bar q$ pairs of different
transverse size, $b$ with the target. However analysis of eq.(1) indicates
that at very small $x$ this equation should break down since
it violates the unitarity limit
which requires that  inelastic  cross  section  for the hard
 interaction  of  a
system of transverse area $S$ cannot exceed $\sigma_{inel} \le S$.
This condition has earliest practical implications in the case of heavy nuclei
where it corresponds to \cite{AFS,FKS}
\begin{equation}
\alpha_s(Q^2)xG_A(x,Q^2) \le {1.9 Q^2 \over A^{1/3}}
\end{equation}
Equation (3) indicates that hard processes sensitive  to gluon densities
in nuclei at small $x$ and moderate $Q^2 \le 10 GeV^2$ should
be strongly shadowed.
For example, in the case of $\sigma_L(A)$ we find that for  $Q^2= 10 GeV^2$
it cannot exceed the value corresponding to $xG^{eff}_A(x,Q^2) \le
{50 \over A^{1/3}}$. Soft gluons  are expected to be shadowed strongly as well
(stronger than soft quarks) \cite{FLS93}.
Therefore very significant shadowing is expected for
$\sigma_L(A)$
for large A at $x \le 10^{-2}$.

An increase of shadowing for $\sigma_T(A,x,Q^2)$ in the limit of small $x $ and
fixed $Q^2$ is also expected due to several effects: a) increase of the soft
cross section with $s$ ($\propto s^n$ with $n \ge 0.08$, b) contribution of
the triple Pomeron term, c) larger shadowing of the small size configurations
due to larger gluon shadowing.

\subsection{Nuclear effects in the diffraction production of vector mesons}

\subsubsection{Coherent production of vector mesons off nuclei  at small t}

The QCD analysis \cite{Brod94} confirms the conjecture of
refs.~\cite{FS88,BM88,KNNZ93} that at large $Q^2$ vector mesons are produced
in small transverse size configurations (at least for the
longitudinally polarized photons) and hence the color transparency
phenomenon (CT) is expected.  In the case of coherent vector meson
production off nuclear targets QCD prediction, in the form of
equation~~(4), with factor $A^2$ substituted by
$\left(G_A(x,Q^2)/G_N(x,Q^2)\right)^2$,
absorbs all the dependence on the atomic
number in the gluon
and sea quark distributions
 of the target. But it is well known
that the evolution of parton distributions with $Q^2$ moves the effect
of nuclear shadowing to smaller $x$.  Thus at small but fixed $x$ and
sufficiently large $Q^2$ the cross section for hard diffractive
processes is expected to fulfill the following relation:
\begin{equation}
\left. {d\sigma^L_{\gamma^*+A \rightarrow X+A}\over dt}\right| _{t=0} =
A^2 \left. {d\sigma^L_{\gamma^*+N \rightarrow X+N}\over dt}\right| _{t=0} \ .
\label{eq:18cc}
\end{equation}
This is the so called color transparency phenomenon which leads to the
validity of the impulse approximation -- the nucleus is transparent
for the projectile and there is no final state interactions. The onset
of CT should occur at moderate $Q^2$ since gluon shadowing disappears
  fast  with increasing $Q^2$ at fixed $x$~\cite{FLS}. If the size of
the configuration is fixed (at large but fixed $Q^2$) but the energy
of the collision increases, shadowing effects should become more and
more important since the gluon shadowing increases with decreasing
$x$~\cite{FS88}, (see figure~3).
Moreover analysis of the unitarity
constraints in Eq.(3) demonstrates that
in the scattering off heavy nuclei screening effects should lead to
very substantial suppression of coherent vector meson
production cross section
$\left. {d\sigma^L_{\gamma^*+A \rightarrow X+A}\over dt}\right|_{t=0} $
for $x \sim 10^{-4}, Q^2 \sim 10 GeV^2$
as compared to the expectation of eq.(\ref{eq:18cc}).
 Thus the use of heavy nucleus beams would allow to observe
nonlinear effects already in the HERA kinematics, while for the case of
 $ep$ collisions  similar effects are likely to be significant for
$x \le 3 \cdot 10^{-5}$ only \cite{AFS} which is beyond the HERA kinematics.
Similar CT behavior is expected for the
production of transversely polarized vector mesons but at
significantly larger $Q^2$ than for the longitudinally polarized
vector mesons, real photons.

As explained above, the preliminary HERA data indicate that PQCD
predictions contained in formula~(1) are applicable already
for $Q^2 \sim 10$~GeV$^2$.  Obviously this is an implicit
confirmation of the color transparency logic since it confirms both
the presence of small transverse configurations in the $\rho$ meson
and the smallness of their interactions with hadrons.  It would be
important to investigate further these effects more directly at ultra
high energies. To this end we consider briefly the scattering off the
lightest nuclei \cite{FMS95}.  Note that there are discussions to
accelerate deuterons at HERA and to polarize them in order to measure
the parton distributions in the neutron.

\subsubsection{Color transparency effects in  $\gamma^*_L
 + D (A) \rightarrow V_L +D(A)$.}


The very existence of the color coherence effects leads to
a rather nontrivial dependence of the
cross sections of hard diffractive processes on $x,Q^2$. To
elucidate this point we consider in this section  diffractive
 electroproduction of vector mesons off the deuteron.

First, let us consider the leading twist effect. It follows from
eq.(\ref{eq:18cc}) that at t=0 the amplitude of this process is proportional to
the parton density in the deuteron. The nuclear effect in the leasing
twist depend on $x,Q^2$ in a rather complicated way. At $x \sim 0.1$
and $Q^2 \sim ~~ few~~ GeV^2$ - the kinematics
of the HERMES facility the gluon density in nuclei is significantly
enhanced:$ {G_A(x,Q^2) \over A G_N(x,Q^2)} > 1$. This effect follows
from the need to reconcile the momentum and baryon sum rules with the
${F_{2A}(x,Q^2) \over F_{2N}(x,Q^2)}$ data \cite{FLS}.
The dynamical mechanism relevant for the gluon
enhancement is not understood so far.

Consequently QCD predicts an enhancement but not shadowing for
the electroproduction of vector mesons at $t=0$ off the deuteron
at $x \sim 0.1$.
This effect
should die out rather rapidly with increase of $Q^2$ due to the QCD evolution
of parton distributions with $Q^2$ (cf. Fig.  for the $Q^2$ dependence
of parton distributions).

At sufficiently small $ x \le 10^{-2}$ shadowing of gluon distribution
dominates.
We will restrict the discussion to the region of sufficiently large
$x \ge 10^{-4}$ where   interaction of a small $q \bar q$ state with a nucleon,
 $\sigma_{q \bar qN}(b^2,x)$ which is given by eq.(1) is small as
compared to the unitarity limit and therefore
 QCD evolution equations seem to be
applicable. In this kinematics one expects a fast decrease
of shadowing with increase of $Q^2$.

Obviously, at $t \approx t_{min}$ shadowing effects are
small since inter-nucleon distance in the deuteron are comparatively
large. To enhance these effects it would be advantageous to study
experimentally the coherent electroproduction of
vector mesons at $|t| \ge 0.5 GeV^2$ where an interesting
diffraction pattern with secondary maximum
 was observed long time ago for photoproduction of $\rho$-meson.
This pattern at $Q^2=0$ arises within the vector dominance model
as a result of the vector meson rescatterings. At large $Q^2$ QCD predicts more
complicated behavior.

Let us consider firstly rescatterings of the produced $q \bar q$ pair of small
size $b$.
In this case
the scattering amplitude is given
by the sum of two terms, one given by the impulse approximation and
the other due to double scattering:
\begin{eqnarray}
{d\sigma_L(\gamma^*+D \rightarrow V+D)\over dt} =\\
 {1 \over 16 \pi}\int \left| 2S_D(t)f_{\gamma^*N \rightarrow VN}(x,b^2,r_t)
+\nonumber \right. \\
\int d^2k_T [{i \over 8 \pi^2} \nonumber \left.
f_{\gamma_L^*N \rightarrow VN}(x,b,Q^2, r_t/2-k_t)\right. \\
 f_{q \bar q, N}(x,b, r_t/2 + k_t)S_D(4k_t^2)]\\
\left.
 \psi_{\gamma^*}(z,b,Q^2)
 \psi_V(z,b) dz d^2b   \right| ^2 \ ,
\label{eq:18c1}
\end{eqnarray}
where $t=-r_t^2$, $S_D(t)$ is the deuteron form factor, and $f_{q \bar
  q, N}(x,b,r_t/2+k_t)$ is the amplitude for the elastic
rescattering of the $q \bar q$ pair.  For simplicity we ignore here the
spin indices.  For the interaction of a small transverse size $q \bar
q$ configuration small impact parameters $b$ dominate in
equation(\ref{eq:18c1}).  Hence the CT prediction of
formula~(4) is that at small but fixed $x$ with increasing
$Q^2$ the relative contribution of the second term should be
proportional to ${1\over Q^2}xG_N(x,Q^2)$.  Since at $-t \ge -t_0 \sim
0.5 {\rm~GeV}^2$ the elastic cross section is dominated by the
square of the second term,
this mechanism leads
in this region to the cross
section which is
extremely sensitive to the CT effects. In particular,
the ratio
\begin{equation}
\left. {d \sigma_L^{\gamma^* + D
\rightarrow V +D}\over dt }\right| _{-t \ge -t_0} /
\left. {d \sigma_L^
{\gamma^* + D   \rightarrow V +D} \over dt }\right| _{-t=0}  =
\label{gammad}
\end{equation}
\[\left| {\left<\sigma_{q\bar qN}(b)\right>
 \over 4 \pi} \left< {1\over R^2} \right> \right| ^2
{\exp Bt \over 4}
  \propto {x^2 G^2_N(x,Q^2) \over Q^4},
\]
where
\begin{equation}
\left<\sigma_{q\bar qN}(b)\right> = {\int d^2b
\psi_{\gamma^*_L}(b) \psi_{V}(b)\sigma^2_{q\bar qN}(b)
\over \int d^2b \psi_{\gamma^*_L}(b) \psi_{V}(b)\sigma_{q\bar qN}(b)}
\label{sigm2}
\end{equation}
should strongly decrease with increasing $Q^2$ and flatten for
sufficiently large $Q^2$ to a leading twist behavior due to the
space-time evolution of the $q \bar q$ configurations. On the contrary
at fixed $Q^2$ this ratio should increase with decreasing $x$. Here
$B=B_{\gamma^* N}/2$ with $B_{\gamma^* N}$ denoting the slope of the
differential cross section for the elementary $\gamma^* + N
\rightarrow V + N$ reaction and $\left<{1\over R^2}\right> =\int d^3r
r^{-2}\psi_D^2(r)$.  The large $t$ ($-t \ge 0.5 {\rm~GeV}^2$) dependence
of the cross section in equation~(\ref{gammad}), ${d \sigma \over dt}
\propto \exp(B't)$ with $B' \sim 2 {\rm~GeV}^{-2}$, is significantly
weaker than in the Glauber model where $B'$ is expected
to be
\begin{equation}
B'={B_{\gamma^* N} B_{\gamma N} \over  B_{\gamma^* N} + B_{\gamma N}}
\approx 3 {\rm~GeV}^{-2 } \ .
\label{slope}
\end{equation}
We neglected here the deuteron quadrupole form factor effects. They
lead to a contribution to the cross section which does not interfere
with the electric transition and for which Glauber effects are small.
This contribution fills the minimum due to the interference of the
impulse and double scattering terms~\cite{Franco}. However this
contribution to the cross section can be significantly suppressed by
using a polarized deuteron target. Similar effects should be present
for the scattering off heavier nuclei, like $^{3,4}He$. The
measurement of the depth of the Glauber minimum due to the
interference of the amplitude given by the impulse approximation with
rescattering amplitudes would allow to check another feature of
expression~(1), namely the large value of the real part of
the production amplitude $Re f/Im f \sim \pi n/2 \sim 0.5$, where $n$
characterizes the rate of increase of the gluon density at small $x$,
$ xG_N(x,Q^2) \propto x^{-n}$.
In this discussion we neglected the leading twist mechanism of
double rescattering related to the leading twist nuclear shadowing.
It is likely to have similar $t-$dependence as the term we considered
above. It may compete with the mechanism we discussed above in
a  certain
$x,Q^2$ range. This question requires further studies. In any case
it is clear that in a wide kinematic range the relative hight of the secondary
maximum would be strongly suppressed as compared to the
the case of the vector meson production by a real photon.
At very small $x$ for $Q^2$ where $\sigma_{q \bar q N}$ is close to
unitarity bound this suppression may disappear. This would establish
the $x,Q^2$ range where color transparency should disappear.

Recent FNAL data on incoherent diffractive electroproduction of vector
mesons off nuclear targets~\cite{FNAL94} did find an increase of
nuclear transparency with increasing $Q^2$ as predicted
in~\cite{FS88,BM88,KNNZ93}. However a significant effect is reported
for a $Q^2$ and $x$ range where the average longitudinal distances are
comparable with the nuclear radius $l_c={1 \over 2 m_N x} \sim R_A$
and it is well known that at large $x$ shadowing disappears for hard
processes. Thus it is necessary to investigate theoretically to what
extent the observed increase of transparency is explained by the
effects of finite longitudinal distances. The ideas discussed in this
report do not apply directly to color transparency phenomena at moderate
energies.  For a recent review of this field we refer the interested
reader to~\cite{FMS94}.

\vspace{2cm}

\end{document}